\documentclass[11pt]{iopart}
\usepackage[cp1251]{inputenc}
\usepackage[english]{babel}
\usepackage{amssymb}
\inputencoding{cp1251}

\usepackage{ifpdf}
\ifpdf
	\usepackage[pdftex]{graphicx}
	\usepackage{epstopdf}		
\else
	\usepackage{epsfig}
	\usepackage[dvips]{graphicx}
\fi
\usepackage{xcolor}

\newcommand{\be}{\begin{eqnarray}\label}
\newcommand{\ee}{\end{eqnarray}}
\newcommand{\bib}{\bibitem}
\newcommand{\prt}{\partial}
\newcommand{\p}{\prime}

\makeatletter
\def\@fnsymbol#1{{}^{\arabic{footnote}}}
\def\@makefnmark{\hbox{$\fnsymbol{footnote}\m@th$}}
\renewcommand\footnoterule{%
  \kern-3\p@
  \hrule\@width.4\columnwidth
  \kern2.6\p@}
\makeatother

\begin{document}

\begin{center}
{\Large \bf Collective Lorentz invariant dynamics \\ on a single ``polynomial'' worldline}
\normalsize
\vskip2mm
{Vladimir V. Kassandrov, Ildus Sh. Khasanov, Nina V. Markova}
\vskip2mm
Institute of Gravitation and Cosmology, \\ 
Peoples' Friendship University of Russia, Moscow, Russia
\end{center}

\begin{abstract}
Consider a worldline of a pointlike particle parametrized by polynomial functions together with the light cone (``retardation'') equation of an inertially moving observer. Then a set of apparent copies, R- or C-particles, defined by the (real or complex conjugate) roots of the retardation equation will be detected by the observer. We prove that  for any ``polynomial''~worldline the induced collective dynamics of R-C particles obeys a whole set of canonical conservation laws (for total momentum, angular momentum and the analogue of  mechanical energy). Explicit formulas for the values of total angular momentum and the analogue of total rest energy (rest mass) are obtained; the latter is ``self-quantized'', i.e. for any worldline takes only integer values.  The dynamics is Lorentz invariant though different from the canonical relativistic mechanics. Asymptotically, at large values of the observer's proper time, the R-C particles couple and then assemble into compact incoming/outgoing clusters. As a whole, the evolution resembles the process of  (either elastic or inelastic) scattering of a beam of composite particles. Throughout the paper the consideration is purely algebraic,  with no resort to differential equations of motion, field equations, etc. 
\end{abstract}

\section{Introduction}

We continue to elaborate the concept of the so-called {\it ``one-electron-Universe''} initiated long ago by J.A. Wheeler and R.  Feynman~\cite{Feynman1,Feynman2}. The question is about the construction of self-consistent dynamics of a set of identical particles that all belong to one and the same worldline. Specifically, in~\cite{Khasan,Vestnik} we considered the class of pointlike particles' worldlines defined {\it implicitly} by a system of algebraic equations
\begin{equation}\label{implicit}
\Psi_a(x,y,z,t)=0,~~~a=1,2,3. 
\end{equation}
where $\{\Psi_a\}$ are three arbitrary independent functions of three spatial coordinates and the timelike parameter $t$. Then, for any value of $t$, one has a finite set of roots $\{x_k,y_k,z_k\},~~k=1,2,\dots,N$ defining the  positions of $N$ pointlike particles. The latter are considered {\it identical} and possess, in particular, equal masses and charges. In the course of time, the particles-roots move along the trajectory curve resulting from (\ref{implicit}) (through elimination of $t$) which, nonetheless, may consist of a number of isolated {\it branches}. Anyway, the fact that all the particles belong to the same ``worldline''  imposes  rigid  restrictions on their dynamics since the roots are strongly correlated. 

These correlations become manifest when the generating functions $\{\Psi_a\}$ in (\ref{implicit}) are arbitrary {\it polynomials}, both in $\{x_a\}$ and $t$. Indeed, after two of three spatial coordinates are eliminated, the system reduces  to a polynomial equation for one variable, with coefficients being some polynomials in $t$. For such a system, 
the above correlations are explicitly represented by the well-known {\it Vieta formulas}. Moreover, for any {\it non-degenerate} (in the sense specified in~\cite{Vestnik}) form of generating polynomials 
these formulas give rise to a system of {\it conservation laws} reproducing standard laws of Newtonian mechanics, namely, the law of conservation of total momentum and (the analogue of) total mechanical energy. Total angular momentum is also conserved for any non-degenerate polynomial system. 

Our main goal in the paper is to make an attempt to put the scheme in correspondence with the requirements of Special Relativity. The realization is undertaken in Sect.2 and 3 where we consider a ``unique worldline'' and an inertially moving observer, together with corresponding  {\it retardation equation} (equation of the observer's light cone) marking a set of points on the worldline. Note that such points, in fact apparent copies of one and the same particle, have  been considered previously in ~ \cite{GinzBolot,Bolot} (for {\it tachionic} worldlines) and in~ \cite{Duplic,YadPhys} (for the case of the {\it complexified} space-time background). 

In the paper we consider only the familiar class of worldlines defined in the canonical parametric way, i.e. by $X_\mu=F_\mu (\tau),~~\mu=0,1,2,3$. Even in this simplest case one obtains, from the principal equation of the light cone of an ``inertial'' observer, an ensemble of identical particles with nontrivial correlated dynamics. Again, the correlations become manifest when the parameterizing functions $\{F_\mu\}$ are arbitrary polynomials in $\tau$. In this case, we immediately get the polynomial equations for each of the coordinates $\{X_\mu\}$ and the parameter $\tau$ itself, with coefficients depending on the (proper) time of the observer $T$. Applying then  the Vieta formulas, one again obtains, {\it for an arbitrary ``polynomial'' worldline}, a full system of conservation laws for the set of roots-particles! 

Note also that one should take into account both real and complex conjugate roots  of the considered polynomial system of equations. The latter have equal real parts and, w.r.t. to the latter, can be visualized in the observable 3-space. On the other hand, both parts of complex conjugate roots enter into the Vieta formulas and, consequently, contribute to the conservation laws. Thus, the complex conjugate roots should be identified with the second, {\it composite} kind of particles (with twice greater mass); in~\cite{Khasan, Vestnik} they were called C-particles (to distinguish with R-particles corresponding to real roots). 
 
The Lorentz invariant~\footnote{In fact, one can exploit the more general {\it conformal} symmetry of the light cone equation} structure of the equation of light cone allows one to regard the whole scheme as a {\it relativistic generalization} of the previously developed Newtonian-like {\it algebraic mechanics}. Then one should consider as physically meaningful only the $SO(3,1)$-covariant characteristics of the dynamical system of R-C particles. This requirement results in a rather specific set of the conservation laws which do not fully reproduce those specific for the canonical relativistic mechanics. 

Moreover, velocities of R-particles are, generally, not limited. Indeed, as in the purely Newtonian case~\cite{Khasan,Vestnik}, at some instants of the observer's time $T$ he observes an event of {\it merging} of a pair of R-particles when some two real roots become {\it multiple} and then transform themselves  into a pair of complex conjugate ones. Physically, this situation models the process of {\it annihilation} of two R-particles (one of which should be considered as an {\it antiparticle}) accompanying by the formation of one composite C-particle. At the points of annihilation, singular points of the worldline, velocities of the colliding R-particles become infinite. However, there exist some possibilities to avoid the difficulty of superluminar velocities. These are discussed throughout the paper,  together with other problems related to the correspondence of the elaborating  scheme with the principles of Special Relativity. 

The organization of the paper is as follows. In Sect.2 we describe the principal system of the light cone equation and that of a ``polynomial'' worldline in ``natural parametrization'' $X_0(\tau)=\tau$ and seek for the properties of its roots-particles. Remarkably, the (conserved) total energy (rest mass) of the system of R-C particles can take only integer values (equal to the higher degree of the parameterizing polynomials). One can also find therein a simple explicit formula that relates the value of the (conserved) total angular momentum to the  leading coefficients of the parameterizing polynomials.  

In Sect.3 we investigate the other, so-called ``timelike'' class of polynomial worldlines, with $p>n$, where $p$ and $n$ are the higher degrees of the polynomials parameterizing the temporal and the spatial coordinates on the worldline, respectively. In doing this, we obtain the manifestly Lorentz invariant form of all three canonical conservation laws and discuss their physical meaning and interrelations with SR. The total energy (rest mass) appears again to be ``self-quantized'' being always equal to the number of roots.  

In Sect.4 we consider the (unexpectedly nontrivial) asymptotic behavior of the ensemble of R-C particles, at large values of the observer's time $T$ and for the class of timelike worldlines (defined by polynomials of a rather great order $p$). We demonstrate that, on the background of general {\it runaway} of particles, there are at least two time scales ($T_1$ and $T_0\gg T_1$) at which peculiar phenomena of ``self-organization'' of the particles' ensemble do take place. Specifically, at $T\simeq T_1$ separate R-C particles start to {\it couple} while at $T\simeq T_0$ the pairs {\it form themselves ``multi-particle'' clusters}. 
Particular details of the process, including its time (ir)-reversibility, depend on the {\it parity} of $p$ and $n$ and their {\it multiplicity}. Finally, we give a graphical illustration of different stages of the process and  speculate on the possible physical interpretation of the particles' evolution (in particular, as a model of (in)elastic scattering). 

Concluding remarks concerning the advantages, drawbacks and perspective  investigations of the ``polynomial mechanics'' are presented in Sect.5. In the Appendix A we prove (analyzing the structure of the so-called Sylvester matrices) explicit formulas for the total rest energy and angular momentum of an arbitrary  system of R-C particles. In the Appendix B we present an illustrative example of collective dynamics of the 48 (R-C) particles on a single polynomial worldline, calculate the exact values of all of the related conserved quantities and trace the most important stages of the evolution.

\section{Conservation laws for polynomial worldlines in natural parametrization}

Consider a  parametrically defined worldline $X_\mu = F_\mu (\tau),~~\mu=0,1,2,3$, with $\tau$ being a monotonically increasing timelike parameter (not necessarily the proper time one!). The worldline itself defines the motion of one and only one pointlike particle. Nonetheless, if we consider the process of ``detection'' of the latter by an ``observer'' moving along its own worldline $\xi_\mu = f_\mu(T)$, we encounter the effect of ``multiplication'' of the initial particle and formation of apparent copies of it located at different points on one and the same worldline. For this, one should make use of the well-known {\it equation of the light cone of the observer} (``the retardation equation'')~\footnote{Note that we do not assume any Minkowski structure \'a priori; velocity of light is set to be unit, $c=1$}:
\be{retard1}
(X_0(\tau) - \xi_0(T))^2 = \sum_{a=1}^3 (X_a(\tau)-\xi_a(T))^2 .
\ee    

Generally, for any fixed position of the observer specified  by the value of $T$, equation (\ref{retard1}) could have a lot of roots $\{\tau_i\}$, including real-valued ones, which define, with respect to the observer, the positions and collective dynamics of particles that all belong to the same worldline.  However, it is well-known (see, e.g.,~\cite{Jackson}) that for the class of worldlines which contain no segments with superluminar velocities this is impossible: equation (\ref{retard1}) has in this case only one root for which $X_0<\xi_0$ (the {\it retarded} solution) and one ``unphysical'' with $X_0>\xi_0$ (the {\it advanced} solution). 

Effect of ``multiplication'' for the class of purely {\it tachionic} worldlines and some of its  interesting consequences were considered in the papers of B.M. Bolotovskii, V.L. Ginzburg and V.P. Bykov ~\cite{GinzBolot,Bolot}. as well as in the works of M. Ibison (see, e.g.,~\cite{Ibison}).  For the case of complex-valued worldlines an analogous construction had been presented in our works~\cite{Duplic, YadPhys}. Remarkable properties of the worldlines on the {\it complexified} space-time background concerning, in particular,  the procedure of generation of the Kerr-type solutions to the Einstein-Maxwell electro-vacuum system of equations, had  been also numerously investigated by E.T. Newman~\cite{Newman1,Newman2}, A.Ya. Burinskii~\cite{Burin1,Burin2} et al. 

Generally, the class of worldlines containing segments with superluminar velocities and forbidden, therefore, in the canonical Special Relativity had been first proposed by E.C.G. Stueckelberg~\cite{Stueckel1,Stueckel2} who used these, in particular, to describe the annihilation/creation processes. Later such worldlines were examined  by J.A. Wheeler and R. Feynman~\cite{Feynman1,Feynman2} in the framework of the so-called ``one-electron-Universe'' paradigm. Of course, many problems arising from the permit of superluminar velocities are still on the agenda. However, there are reassuring expectations to avoid these difficulties: one of these will be considered in Sect.4. 

Our main goal here and below is to demonstrate the following remarkable fact.  For any worldline $X_\mu(\tau)$ for which the parameterizing functions $F_\mu(\tau)$ are arbitrary polynomials, corresponding set of roots-particles fixed by an ``inertial'' observer, obeys a number of conservation laws. The latter are defined for a set of $SO(3,1)$-covariant quantities related to the roots of (\ref{retard1}) and their time derivatives. The induced conservative dynamics is Lorentz invariant though different from the canonical relativistic one~\footnote{For example, the field degrees of freedom do not contribute explicitly to the conservation laws; we suspect that these are indirectly represented in the characteristics of particles themselves, see the discussion below}.      

Consider now the defining equations of an arbitrary ``polynomial'' worldline $X_\mu(\tau)$ together with the light cone equation for an inertially moving observer $\xi_\mu = \xi_\mu^{(0)} + v_\mu T$. For simplicity, let us  first accept the natural parametrization of the worldline $X_0(\tau)=\tau$ and, as for the observer, assume him to be resting at the origin, $\xi_0=T,~~\xi_a =0$. Then (\ref{retard1}) takes the form
\be{retard}
F:=X(\tau)^2 +Y(\tau)^2 +Z(\tau)^2 - (T-\tau)^2=0,
\ee
in which
\be{polynomsXi}
\left\{ \begin{array}{lll}
X_1 \equiv X(\tau)=a_x\tau^n+b_x\tau^{n-1}+\dots +e_x,\\ 
X_2 \equiv Y(\tau)=a_y\tau^n+b_y\tau^{n-1}+\dots+e_y,\\
X_3 \equiv Z(\tau)=a_z\tau^n+b_z\tau^{n-1}+\dots+e_z,
\end{array}\right.
\ee
are arbitrary polynomials of a highest degree $n$ in $\tau$ (some of the three polynomials may be of smaller degrees). Below we assume $n\ge 2$ and note that  the leading term of degree  $2n$ in (\ref{retard}) does not ever cancel. 

Then, at any instant of the observer's time $T$, (\ref{retard}) has exactly $2n$ roots some of which are real while others enter in complex conjugate pairs. Corresponding copies-particles will be called R- or C-ones, respectively. The latter, as it was explained in the introduction, are located off the real worldline points but can be nevertheless visualized according to equal real parts of complex conjugate roots. 

Specifically, polynomial equation (\ref{retard}) acquires the following structure:
\be{terms}
F=A \tau^{2n} +B\tau^{2n-1}+C \tau^{2n-2}+\dots + (D +2 T)\tau + (E - T^2) = 0, 
\ee
 where the coefficients $A,B,C,\dots,D,E$ do not depend on the time $T$ at all; in particular, 
\begin{equation}\label{coeff3}
\begin{array}{l}
\fl A=a_x^2+a_y^2+a_z^2\equiv \vert \vec a \vert^2, ~~B=2(a_xb_x+a_yb_y+a_zb_z)\equiv 2\vec a\cdot \vec b, \dots, \\
\fl E=e_x^2+e_y^2+e_z^2\equiv \vert \vec e \vert^2. 
\end{array}
\end{equation}
Now, making use of the first, linear in roots {\it Vieta formula} applied to the polynomial equation (\ref{terms}), we immediately obtain   
\be{sumtau}
\sum \tau_i = - B/A =constant,
\ee 
where summation here and below runs over all the roots $i=1,2,\dots,2n$. Thus, we obtain the first  ``conservation law'' for the quantity which is usually associated with the so-called ``retarded time''. However, the values $\{\tau_i\}$ of the particles' timelike parameter $\tau$ may here be complex, decrease in the course of the monotonic increasing of the observer's time $T$ (then the corresponding root represents an {\it antiparticle}, see Sect.4), etc. In fact, only the latter ``macro-time'' is a well-defined {\it evolution parameter} of the theory. 

Consequently, the derivative quantity $\dot \tau:=d\tau/dT$ will be conserved as well, 
\be{sumtau1}
\sum_i \dot \tau_i = 0, 
\ee
while the next, quadratic in roots Vieta-like formula~\footnote{Precisely, we make use of the formula for the {\it sum of squares of the roots} which can be obtained from the quadratic Vieta formula with account of the first, linear in roots one} and its derivatives  w.r.t. $T$ lead to the following relations:
\be{sumtau2}
\sum \tau_i^2 = (B/A)^2 - 2C/A = constant, ~~\sum \tau_i \dot \tau_i =0,~~\sum \dot \tau_i^2 + \ddot \tau_i \tau_i = 0.
\ee

Consider now one of polynomial equations for spatial coordinates of R-C particles, say  for $x\equiv X_1$:
\be{coord}
H_x:=x-X(\tau)=x-(a_x\tau^n+b_x\tau^{n-1}+\dots +e_x) =0. 
\ee
Together with (\ref{retard}) it forms a joint system for determination of the implicit dependence of $x$ on $T$; to do this, one should eliminate $\tau$ by taking  corresponding {\it resultant} $D_x:=Res[F,H_x,\tau]$ and equating it to zero. Analyzing the structure of the determinant $D_x$ of the arising  {\it Sylvester matrix} (see Appendix A for details), one obtains  that the leading terms in the resultant equation will have the following structure: 
\be{result1}
D_x= P x^{2n} + Q_x x^{2n-1}+ (R_x+E_x T+ G_x T^2)x^{2n-2}+\dots = 0,
\ee
where $P,Q_x,R_x,E_x,G_x$ are time-independent constants; in particular, we are interested below in the following two coefficients:
\be{coeffX}
P=A^n=\vert \vec a \vert^{2n}, ~~~G_x=-na_x^2 A^{n-1}\equiv -na_x^2 \vert \vec a\vert^{2n-2}   
\ee
in which the quantity $A$ was defined in (\ref{coeff3}). 
Quite analogous expressions for two other coordinates one will obviously come to.  

Now, making use again of the first two Vieta-like formulas, one obtains the following constraints between spatial coordinates of particles:
\be{duplcoord1}
\sum x_i = -\frac{Q_x}{P} = const,
\ee
\be{duplcoord2}
\sum x_i^2 = (\frac{Q_x}{P})^2 - 2\frac{R_x + E_x T + G_x T^2}{P},
\ee
and similar formulas for $y$- and $z$- coordinates of particles. From the linear constraints like (\ref{duplcoord1}) we see that the selected reference frame of the observer is just the center-of-mass one (on account of the assumed {\it identity} of R-particles which, therefore, possess equal masses) so that the {\it total momentum is conserved and equal to zero}:
\be{Moment}
\sum (\dot x_a)_i \equiv  \sum (x_a^\p)_i \dot \tau_i = 0,~~~a=1,2,3 
\ee
where $\p$ denotes differentiation w.r.t. $\tau$. 

Remarkably, conservation laws (\ref{duplcoord1},\ref{duplcoord2}) (as well as others below examined) are valid only for an {\it inertially moving} observer (including the currently considered case of the observer {\it at rest}). For {\it nonlinear} polynomial worldlines of an observer (corresponding to a {\it non-inertial} reference frame) the structure of resultants like (\ref{result1}) becomes more complicated, and {\it conservation laws turn out to be broken}.   

Composing now the $SO(3)$-invariant combination from the sums of squares like  (\ref{duplcoord2}) and differentiating twice the result  by $T$, one obtains the following conservation law for the ensemble of R-C particles:
\be{conservsquare}
W:=\sum ({\dot x}_a)_i ({\dot x}_a)_i+ \sum ({\ddot x}_a)_i (x_a)_i   =-2 \frac{G_x+G_y+G_z}{P}=   constant.  
\ee
Substituting then expressions  (\ref{coeffX}) for the constants $P,G_x$ (and for  corresponding ones defining  $G_y,G_z$) we obtain finally
\be{QuantRule}
W=2n\frac{a_x^2+a_y^2+a_z^2}{A}\equiv 2n.
\ee
Thus, we have obtained a sort of {\it quantization rule} for the quantity which, in fact, is the analogue of {\it total energy} of the R-C ensemble:
\be{VirQuant}
\frac{W}{2}=K+V = const = n , 
\ee
where the first term $K:=\frac{1}{2}\sum ({\dot x}_a)_i ({\dot x}_a)_i $ reproduces the {\it total kinetic energy} while the second one $V:=\frac{1}{2}\sum ({\ddot x}_a)_i (x_a)_i$ stands for the quantity known in classical mechanics as {\it virial} and proportional (for radial forces homogeneously depending on mutual distances) to the {\it total potential energy} (see, e.g.,~\cite{Goldstein,LandauMech}).  Consequently, {\it any ``polynomial worldline'' (in natural parametrization $X_0(\tau)=\tau$) of a highest degree $n$ defines a conserved energy-like $SO(3)$-scalar quantity which is moreover universal, integer and equal to the degree $n$}.

Let us demonstrate further that the vector of total angular momentum 
\be{angmom2}
\vec M = \sum \dot{\vec r} \times \vec r
\ee
is also conserved for any R-C system of particles. To do this, implicitly differentiating the principal equation (\ref{retard}),  we come to the {\it rational} expression for $\dot \tau$ (in which  
$\dot F:= \prt F/\prt T,~~F^\p:= \prt F/\prt \tau$):
\be{dertau}
\dot \tau = -\frac{\dot F}{F^\p}= \frac{T-\tau}{(T-\tau)+X X^\p+Y Y^\p+Z Z^\p}
\ee
and then obtain for a component of angular momentum of a particle, say for $M_x$,  expression of the form
\be{componang}
M_x =  Y\dot Z - Z\dot Y = (Y Z^\p - Z Y^\p)\dot \tau. 
\ee 
On account of (\ref{dertau}), expression (\ref{componang}) reduces (except in the singular points of mergings) to the following  equation:
\begin{equation}\label{polynomang}
\begin{array}{l}
\fl M_x F^\p +(Y Z^\p - Z Y^\p) \dot F=0~ \Leftrightarrow~ \\
\fl N_x:= M_x ((T-\tau)+XX^\p+YY^\p+ZZ^\p) - (T-\tau)(Y Z^\p - Z Y^\p)= 0,
\end{array}
\end{equation}
in which the polynomial $N_x(\tau,T,M_x)$ is of the degree $2n-1$ in $\tau$ (and linear in $T$ and $M_x$).  Eliminating then $\tau$ via taking the resultant $R_x:=Res[F,N_x,\tau]$ and equating it to zero, we come to a  polynomial equation $R_x(T,M_x)=0$ of the expected degree $2n$ in $M_x$. The latter  implicitly defines the dependence of (the first component of) angular momentum of particles on the observer's time $T$. 

By virtue of the specific structure of the determinant of the corresponding  Sylvester matrix (see the Appendix A), we obtain that the two leading coefficients in the resultant equation are of the following appropriate form:  
\be{angVieta}
R_x = \Delta(T) M_x^{2n} + \Delta(T) \alpha_x M_x^{2n-1}+\dots = 0,  
\ee     
where $\alpha_x$ is a time-independent constant (see the Appendix A) while $\Delta(T)$ is a specific polynomial in $T$, the {\it discriminant} of the principal equation (\ref{retard}), which turns to zero only at singular points of mergings. The above obtained  structure of resultant (\ref{angVieta}),      
together with the first, linear in roots Vieta formula, guarantees the conservation of total quantity $M_x$: 
\be{angconserv2}
\sum_i (M_x)_i = -\alpha_x = const.  
\ee
The same, of course, is valid for two other components of the vector of total angular momentum $\vec M$. 

To corroborate the property of conservation, we derive in the Appendix A the following universal {\it explicit formula} for the value of $\vec M$:
\be{explicang}
\vec M = 2 \frac{ \vec a \times \vec b}{\vert \vec a\vert ^2},
\ee
which turns out to depend only on the leading coefficients $\vec a=\{a_x,a_y,a_z\},~ \vec b =\{b_x,b_y,b_z\}$ of respective degrees $n$ and $n-1$ in the defining polynomials $X_a(\tau),~a=1,2,3$. If only one polynomial, say $X_1=X(\tau)$, is of a degree $n$ while the two others have lower degrees, the total angular momentum is identically zero.     

Finally, we announce that for the more general parametrization $X_0(\tau)=S(\tau) \ne \tau$ for which the degree $p$ of the polynomial $S(\tau)$ is smaller than $n$, that is $0<p<n$ (the so-called ``spacelike'' worldlines, see below), the above conservation laws all remain valid. However, the proof of this statement as well as corresponding explicit formulas for the total angular momentum (\ref{explicang}) and the analogue of total energy (\ref{QuantRule}) become much more complicated and depend on the particular degrees of generating polynomials (\ref{polynomsXi}). 

\section{``Timelike'' polynomial worldlines and Lorentz invariant dynamics}

The principal retardation equation (\ref{retard}), together with the defining equations of a  polynomial worldline (\ref{polynomsXi}), compose by themselves a Lorentz-invariant system. On the other hand, we recall that only for an {\it inertial worldline of the observer} $\xi_\mu(\tau)$ the conservation laws for the system of R-C particles  remain valid. In that way, we encounter a remarkable 
 relation between the (quadratic) structure of the light cone equation and the conservative character of the collective dynamics it induces in an inertial reference frame.  

Specifically, under a {\it boost}, the parameters $\tau$ and $T$ in (\ref{retard}) should be considered invariant while the four coordinates of both the observer $\xi_\mu(T)$ and particles $X_\mu(\tau)$ transform in the canonical way. This means that in a new inertial reference frame the above used simplest gauge $X_0(\tau)=\tau$ will 
be broken, and the explicit formulas for conservative quantities like (\ref{QuantRule}) or (\ref{explicang}) will be no longer valid. 

Thus, only the $SO(3,1)$-covariant combinations of conservative quantities should be considered as physical meaningful. Such combinations, on the other hand, can be constructed only from the constraints related to the linear or quadratic in roots Vieta-like formulas~\footnote{E.g., formulas for the sums of {\it cubes} of the roots or their time derivatives do not possess well-defined tensor  properties}. As for the first one, the corresponding quantities ${\dot X}_\mu =\{ {\dot X}_0,~{\dot X}_a\}$ evidently form a 4-vector. Since its spatial part, on account of equal (unit) masses of the roots-particles, had been identified with 
the {\it momentum}, the 4-vector ${\dot X}_\mu$ as a whole should be treated as the {\it energy-momentum 4-vector} of an individual particle. 

However, for the ``observer at rest'' in the chosen gauge $X_0(\tau)=\tau$ from Eqs. (\ref{sumtau1}) and (\ref{Moment}) it  immediately follows that the {\it total energy} as well as the {\it total momentum} of the whole R-C ensemble is identically null; this, of course, will then be true in any inertial reference frame and cause serious difficulties concerning the identification of R-C particles in the case of the above used parametrization $X_0(\tau)=\tau$ and, generally, in arbitrary case of the ``spacelike'' worldlines (i.e., for $X_0=S(\tau)$, with $deg(S)=n<p$). Fortunately, the situation changes drastically for the class of ``timelike'' polynomial worldlines which we are going to examine below. 

Specifically, let the parameterizing polynomials of the worldline under consideration have the form (cf. (\ref{polynomsXi})):
\be{timelike}
\left\{ \begin{array}{llll}
X_0=S(\tau)=a_s\tau^p+b_s\tau^{p-1}+\dots+e_s,\\
X_1=X(\tau)=a_x\tau^n+b_x\tau^{n-1}+\dots +e_x,\\ 
X_2=Y(\tau)=a_y\tau^n+b_y\tau^{n-1}+\dots+e_y,\\
X_3=Z(\tau)=a_z\tau^n+b_z\tau^{n-1}+\dots+e_z,
\end{array}\right.
\ee 
where we assume $p>n \ge 2$ and $a_s \ne 0$. Then, for sufficiently large values of $\tau$, corresponding values of the (proper) Minkowski intervals of R-particles $ds^2=dX_0^2-dX_1^2-dX_2^2-dX_3^2\simeq p^2a_s^2 \tau^{2p-2}d\tau^2$ will be evidently {\it positive definite}; this justifies the terms ``timelike'' or, vice versa, ``spacelike''  (precisely, {\it asymptotically timelike/spacelike}) for the characterization of the two different classes of polynomial worldlines. Needless to say that for timelike worldlines the natural parametrization $X_0(\tau)=\tau$ is  forbidden; instead, the ``polynomial reparametrization'' $\tau=A_j \tilde \tau^j +B_{j-1} \tilde \tau^{j-1}+\dots$ is allowed that preserves the property of the initial worldline to be (asymptotically) timelike.

For a timelike worldline (\ref{timelike}), the principal equation of the light cone (\ref{retard1}), for an observer at rest, acquires the form (cf.  (\ref{retard})):
\be{timelikecone}
F:=(T-S(\tau))^2-X(\tau)^2-Y(\tau)^2-Z(\tau)^2=0, 
\ee
so that the polynomial $F$ is of the degree $2p$ in $\tau$, and at any $T$ the observer detects $2p$ associated R-C particles. It is easy to see that all the conservation constraints like (\ref{sumtau} -- \ref{sumtau2}),  for the roots $\tau_i,~~i=1,2,\dots,2p$ or their time derivatives hold equally in the timelike case. As regards for the corresponding timelike coordinates $s_i$ of individual particles,  one can make use of the auxiliary equation 
\be{EqForS}
G:=s-S(\tau)=0 
\ee
to simplify the equation of the light cone (\ref{timelikecone}) to the following form:
\be{timelikecone2}
F= Q^2-X^2-Y^2-Z^2 = 0, 
\ee
in which the quantity $Q:=s-T$ does not contain the parameter $\tau$ so that  (\ref{timelikecone2}) is now of a smaller degree $n$ in $\tau$. Now the structure of resultant of two polynomials (\ref{EqForS}) and (\ref{timelikecone2}), that is of $R_s:=Res[G,F,\tau]$, becomes quite transparent, and after eliminating $\tau$ one obtains the resultant equation $R_s=0$ in the following form:
\be{4energy1}
R_s=Q^{2p} + \alpha(s) Q^{2p-2}+\beta(s) Q^{2p-4} +\dots =0, ~~~(Q=s-T),
\ee  
where $\alpha,\beta,\dots$ are some polynomials in $s$ of degrees {\it at most} $1,2,\dots$, respectively. 
Collecting then in (\ref{4energy1}) the leading terms in $s$ of {\it highest degrees} in $T$ which are, in fact, contained only in the first term $Q^{2p}$ , one obtains 
\be{4energy2}
R_s=s^{2p}-2pT s^{2p-1}+\frac{2p(2p-1)}{2} T^2 s^{2p-2}+\dots =0
\ee 
(note that all other terms do not enter into the principal characteristics written out below and can be discarded). 

Making now use of the corresponding Vieta-like formulas and taking their time derivatives, one obtains from (\ref{4energy2}), in full analogy with the previously described procedure (Sect.2):
\be{4-energy-conserv}
E:=\sum {\dot s}_i = 2p =constant, 
\ee
\be{0-virial}
W_0:=\sum {\dot s}_i^2 +\sum {\ddot s}_i s_i = (2p)^2 - 2\frac{2p(2p-1)}{2}\equiv 2p =const.
\ee

As to the total momentum $\vec P =\{P_a\},~a=1,2,3$ and conjugate  conserved quantities $\{W_a\}$ related to the quadratic in roots Vieta-like  formulas, following the same procedure one immediately concludes  that these both are null in the timelike case, 
\be{timelikemomentum}
P_a := \sum ({\dot x}_a)_i = 0,
\ee
and
\be{a-virial}
W_a:=\sum ({\dot x}_a)_i^2 +\sum ({\ddot x}_a)_i (x_a)_i = 0. 
\ee
We see therefore that, in the timelike case, identification of the quantity  $P_\mu:={\dot X}_\mu =\{ {\dot X}_0,~{\dot X}_a\}$ with the {\it energy-momentum 4-vector} becomes quite adequate. Indeed, the observer {\it at rest} (when the reference frame coincides with the 
{\it center-of-mass frame} of the R-C particles' ensemble) will get the {\it positive-definite total energy} $E$ which, according to (\ref{4-energy-conserv}), is moreover {\it integer, universal and equal to the number of roots-particles $2p$!} One can thus introduce the $SO(3,1$)-invariant 
\be{restmass}
M^2:=E^2-{\vec P}^2=(\sum {\dot s}_i)^2 -  \sum ({\dot x}_a)_i \sum ({\dot x}_a)_i = 4p^2\ge 0, 
\ee
which should be evidently identified with the square of total {\it rest mass} of the particles' ensemble. 

Quite analogously, from the conserved quantities (\ref{0-virial}),(\ref{a-virial}) one can construct the $SO(3,1)$-scalar 
\be{4-virial}
W:=(W_0 - W_1 - W_2 - W_3) = 2p \equiv M\ge 0
\ee
which turns out to be precisely equal to the positive-definite total rest mass of the R-C system! Note that in the 2D non-relativistic case considered in ~\cite{Khasan,Vestnik} corresponding $SO(3)$-invariant quantity conserved by virtue of the quadratic Vieta-like formulas also represents the analogue of total mechanical energy of the system; in the Lorentz invariant dynamics under consideration both ``definitions'' of the total rest energy (rest mass) do coincide, and we regard this fact as  indicative of the {\it self-consistency} of the theory.  

On the other hand, the structure of quadratic relations like (\ref{a-virial}) demonstrates that the ordinary {\it  kinetic-like} part of the total energy represented by the first term therein would be constant only under account of the second term,  the (relativistic analogue of) {\it potential-like} part. The latter, however, directly depends on {\it accelerations} of particles and, in a sense, substitutes for the {\it energy of the fields} associated with particles in the canonical approach. We shall return to discuss these issues in the conclusion.

It is also noteworthy in this connection that in the rest frame the equality of two quantities (\ref{4-energy-conserv}) and (\ref{0-virial}) follows, in fact, from the following identity for the roots of an {\it arbitrary timelike system of equations} (\ref{timelike}),(\ref{timelikecone}) and their time derivatives
\be{identity}
\sum {\dot s}_i \equiv \sum {\dot s}_i^2 + \sum {\ddot s}_i s_i.
\ee
Deep physical sense (if any) and mathematical origins of this remarkable identity related evidently to the total rest energy (rest mass) of the R-C system is not yet revealed.

To conclude, let us write out the manifestly Lorentz invariant form of the conservation laws arising in the scheme under consideration. For the energy-momentum 4-vector one gets
\be{ENMOMINV} 
\sum_{i} \frac{dX_{\mu(i)}}{dT} =C_\mu =const, 
\ee
while for the scalar of total mass one obtains
\be{RESTMASS}
\sum_i \frac{dX_{\mu(i)}}{dT}\frac{dX^\mu_{(i)}}{dT} +\sum_i X_{\mu(i)}\frac{ d^2 X^\mu_{(i)}}{dT^2}= const. 
\ee  
Note that, instead of the individual  proper times of the constituent particles in the canonical approach, one deals here with the unique Lorentz invariant proper time of the (inertially moving) observer $T$. 

Let us finally clarify the situation with the total angular momentum. Recall that in a relativistic invariant scheme one should consider the {\it skew symmetric tensor of angular momentum} with six independent components 
\be{tensorang}
M_{[\mu\nu]} = \sum X_{[\mu} U_{\nu]}, 
\ee
where the quantities $U_\mu$ are the 4-velocities of particles. However, for the considered R-C system, the latter should be defined via differentiations w.r.t. the observer's proper time $T$, that is 
\be{4-velocit}
U_\mu:= \frac{dX_\mu}{dT}\equiv {\dot X}_\mu.
\ee
Three spatial components $M_a:=\frac{1}{2}\varepsilon_{abc} M_{[bc]}$ of the tensor (\ref{tensorang}) evidently reproduce the ordinary expression for the angular momentum vector 
\be{vectangmom}
M_a = \sum \varepsilon_{abc}X_b {\dot X}_c, 
\ee
while  the  components $M_{[0a]}$ are their relativistic counterparts. In our scheme, {\it all these six quantities are conserved for the system of R-C particles as a whole} though, in the timelike case, we were unable to find explicit formulas for them resembling (\ref{explicang}). The $SO(3,1)$-invariant corresponding to the conserved tensor (\ref{tensorang}) is evidently its 4-square $M_{[\mu\nu]}M^{[\mu\nu]}$. Contrary to the square of the energy-momentum 4-vector $P_\mu={\dot X}_\mu$ the latter does not,  however, take a universal value. 

\vskip2mm
\noindent
\section {Asymptotic behavior: formation of clusters \& scattering pattern}

Collective dynamics of R-C particles, in the timelike case, reveals universal remarkable properties at  large values of the observer's time $T$. In fact,  asymptotically, individual particles join into compact groups,  {\it clusters}, which contain two or four R-particles and an essential number of C-particles each. At the same time, different clusters move off from the origin with decreasing velocities. Specific details depend on the particular values of the degrees $p$ and $n$ of the generating polynomials (\ref{timelike}), and we briefly examine them below. 

At sufficiently large values of the observer's time $T$,  the terms of highest  degree $2p$ dominate in the light cone equation (\ref{timelikecone}) parametrized by (\ref{timelike}). Therefore, equation   (\ref{timelikecone}) asymptotically takes the obvious form:  
\be{approxcone}
(T-a_s \tau^p)^2 \simeq 0, 
\ee
with the following, {\it twice degenerate} set  of solutions (to be concrete, let us first assume $a_s>0$ and $T>0$):
\be{asymtau}
\tau_k \simeq \vert T/a_s\vert ^{1/p} \exp{(2\pi \imath k/p)},~~~~k=1,2,\dots,p,\dots,2p . 
\ee
For the limiting values of the (real parts of) coordinates of particles we obtain then from (\ref{timelike}):  
\be{asymXi}
\{X_c\}_k \simeq \Re\{a_c \tau_k^n\} = a_c  \vert T/a_s \vert^{n/p} \cos{(2\pi  k n/p)}, ~~~c=1,2,3. 
\ee
(here and below we rename $\{a_x,a_y,a_z\}\equiv \{a_c\}=\{a_1,a_2,a_3\}$ and so on for $\{b_c\},\dots$). 
However, if for some values of the angle $\varphi_k:= 2\pi k (n/p)$  the coordinates  $\{X_c\}_k$ turn to zero, one should 
take into account the next terms of the degree $n-1$, so that the particles are located much closer to the origin and move slower, respectively. We can call such set of particles {\it exceptional}. 

The asymptotic form (\ref{approxcone}) of the principal light cone equation and corresponding expressions (\ref{asymXi}) for coordinates of particles hold for $T\gg T_0,~~T_0=a_s (b_s/a_s)^p$. The first order corrections due to the account of the terms of the degree $2p-1$ lead, instead of (\ref{approxcone}), to the following approximate form of the light cone equation:
\be{approxcone2}
(T-a_s \tau^p-b_s \tau^{p-1})^2 \simeq 0, 
\ee
valid at the smaller time scale $T_0 > T \simeq T_1$, where $T_1=a_s(\vert \vec a \vert/a_s)^{p/(p-n)}$. Finally, the second order corrections take into account the highest order terms (of degree $n$) of the  polynomials $\{X_c\}$ and lead to the following representation of the principal equation:
\be{approxcone3} 	
(T-a_s \tau^p-b_s \tau^{p-1})^2 - \vert \vec a \vert^2 \tau^{2n} \simeq 0,~~~ \vert \vec a \vert^2: =a_1^2+a_2^2+a_3^2,
\ee
which is valid at $T\simeq T_1$  and breaks the degenerate structure of the roots-particles related to the full square in the right hand side of  (\ref{approxcone2}) or  (\ref{approxcone}). 

It seems quite natural (and most interesting from the physical viewpoint) to examine the case when the whole number of roots-particles $N=2p$ is great, $p\gg 1$. As for $n$, it defines, as a rule, the number of clusters and the number of particles in a cluster which, evidently, are in an inverse proportion to each other.  
Elementary analysis of the successive approximations  (\ref{approxcone3}), (\ref{approxcone2}) and (\ref{approxcone}) of the exact light cone equation (\ref{timelikecone}) leads then to the following conclusions.   

\begin{figure}[ht]
\begin{center}    
    \includegraphics[width=0.8\textwidth]{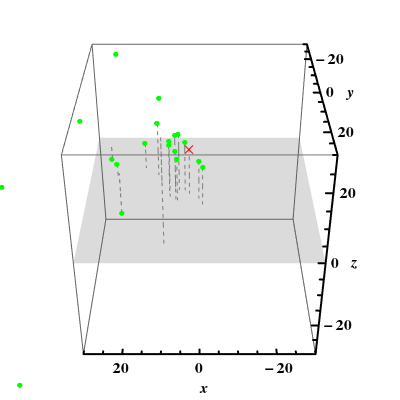}
    \caption{First stage of evolution, $T = 10$,~$T \ll T_1$: R- (in red crosshairs) and 44 C-particles (in green points), some of these being off the depiction space; the pairs are not yet formed}
   	\label{fig:image1}

\end{center}  
\end{figure}

\begin{figure}[ht]
\begin{center}    
    \includegraphics[width=0.8\textwidth]{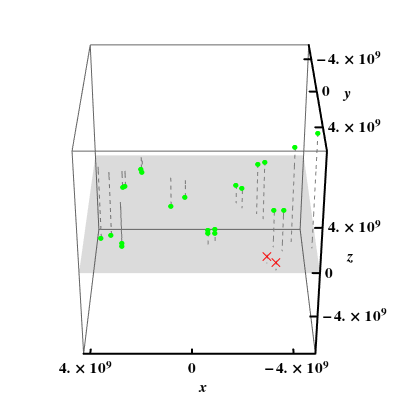}
    \caption{Second stage of evolution, $T=10^{11}$,~$T_0>T\gg T_1$: one R-pair (the other one is far off the frame) and 22 C-pairs; all the pairs are well distinguishable and start to gather into clusters.}
   	\label{fig:image2}
\end{center}  
\end{figure}

\begin{figure}[ht]
\begin{center}    
    \includegraphics[width=0.8\textwidth]{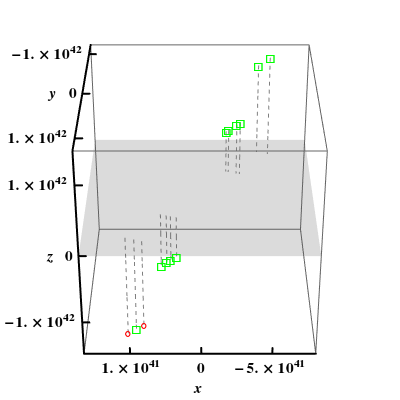}
    \caption{Third stage of evolution, $T = 10^{50}$,~$T\gg T_0$: four well-formed clusters run away along a distinguished 3-direction; two R-pairs are designated by red circles and 22 C-pairs by green squares; individual  particles constituting pairs are too close to be seen.}
   	\label{fig:image3}
\end{center}  
\end{figure}


\vskip2mm
\noindent
{\bf A. At the interval $ 0 \le T \le T_1$} .
\vskip1mm
1.  At particular instants of $T$, some two of R-particles merge (annihilate) and transform themselves into a composite C-particle or vice versa. Before an annihilation event, the course of ``retarded'' time $\tau$ reproduces that of the observer's time $T$ for one of the merging particles and {\it is inverse} for the other one. Thus, one can regard the first/second formation as a particle/antiparticle, respectively. 

2. At $T\simeq T_1$ all the ``events'' are over and, in what follows, the whole number of R- and C-particles remains invariable. Specifically, there are finally only two (in the case of odd $p$) or four (for even $p$)  R-particles. The number of C-particles may be thus very large. This early stage of evolution for the case $p=24, n=20, (N=2p=48)$ is depicted at figure \ref{fig:image1}.  

\vskip2mm
\noindent
{\bf B. At the interval $ T_1 \ll T \le T_0$} .
\vskip1mm
3. All R-particles break up into one (for odd $p$) or two (for even $p$) contracting pairs (R-pairs) while all C-particles form C-pairs consisting of two slightly separated and approaching each other pairs of complex conjugate roots (that is, of four roots-particles altogether). Thus, one observes the phenomenon which can be called {\it the  ``first phase transition''} (see the illustration for the case $p=24, n=20$ at figure \ref{fig:image2}). 
 
\vskip2mm
\noindent
{\bf C. At the interval $ T \gg T_0$ }.
\vskip1mm
4. All (non-exceptional) particles-roots concentrate in vicinity of the straight line defined by the ratio of the coefficients $a_x:a_y:a_z$ and undergo a {\it runaway} according to the law $\propto T^{n/p}$, with decreasing velocities and negative accelerations (note, however, that for $p\gg n$ the motion is nearly uniform). As to exceptional C-pairs, they move along the other direction specified by the ratio of coefficients $b_x:b_y:b_z$; their recession law is $\propto T^{(n-1)/p}$ so that they are located much closer to the origin and move slower, respectively.    

5. For multiple $p$ and $n$, formation of {\it clusters} takes place ({\it the ``second phase transition''}); one of these consists of either one  (for odd $p$) or two (for even $p$) R-pairs ``dressed'' by a number of closely disposed C-pairs. All other clusters consist solely of C-pairs. For $p=24,n=20$ this last stage of evolution is depicted  at figure \ref{fig:image3}~\footnote{In the considered example there are no exceptional particles at all and, thus, only one spatial direction of runaway well noticeable at figure \ref{fig:image3}}. The whole number of clusters, of pairs (as well as particular configurations of the latter within a cluster) depend on the {\it parity} and {\it multiplicity} of $p$ and $n$. In the course of time, separation of clusters evidently grows whereas mutual distances between pairs within a cluster (as well as between particles constituting any pair) {\it relatively} (i.e., in comparison with distances between clusters) decrease.

\vskip2mm
\noindent
{\bf D. ``Full history'' at the whole interval $-\infty >T > +\infty$ }.
\vskip1mm
Making use of the same considerations, one can analogously describe the R-C dynamics on the negative semi-interval and relate, in particular, the {\it incoming} and {\it outgoing} distributions of particles. As a result, one comes to the following conclusions. 
\vskip1mm
6. For odd $p$, the asymptotic equations (\ref{approxcone}) and (\ref{asymXi})  are invariant under the substitution $T\mapsto -T,~~\tau\mapsto -\tau,~~X_c \mapsto \pm X_c$, for even/odd $n$, respectively. Therefore, one has an ``asymptotically time-symmetric''~process resembling the picture of {\it elastic scattering} of a beam of clusters in the transmission/reflection regimes, respectively.

7. For even $p$, equation (\ref{approxcone}) possesses the symmetry $T\mapsto -T,~~ \tau \mapsto \pm \imath \tau,~~X_c \mapsto -X_c$ (for even $n$ only). In particular, for $-\infty<T\ll -T_0$ it has no real roots (so that only C-clusters are present in the incoming  beam) while at $+\infty>T\gg T_0$ it possesses 4 real roots corresponding to a single R-cluster (either ``dressed''  or ``naked'', see also item 5). Thus, the process is {\it ``asymptotically time-asymmetric''} and model that of an {\it inelastic scattering}. At small values of time $T$ one encounters a cascade of creations of R-pairs which successively annihilate then into composite C-particles; the final stage of the process (with one ``dressed'' R-cluster and a lot of C-clusters) was described above, see again the illustration at figure \ref{fig:image3}.

\section{Conclusion}

In fact, we have demonstrated that a quite nontrivial and well defined collective dynamics of pointlike particles can be induced solely by generating functions defining a single  worldline.  For all this, we did not make use of Lagrange functions, systems of differential equations or any other ingredients of canonical mechanics. When the worldline is defined in an ordinary parametric form, an apparent set of particles arises as the set of roots of the equation of light cone of an observer moving along its own worldline. 

We investigated the simplest case of a ``polynomially parametrized'' worldline. Then it is necessary to consider two different kinds of copies-particles, R- or C-ones, that correspond to real or complex conjugate roots of the light cone equation, respectively. Then all physical characteristics should be taken as functions of the proper time $T$ of the observer. 

Remarkably, in the reference frame of an {\it inertially} moving observer (and only in that) the induced collective dynamics turns out to be {\it conservative}, that is, for {\it arbitrary} polynomial worldline obeys a number of {\it conservation laws}. Their structure is completely governed by the {\it Vieta formulas} for the roots-particles and does not include thus any characteristics of the {\it fields} produced by the particles but only the particles' characteristics themselves. On the other hand,  the presented scheme is based on the effect of {\it retardation} and is manifestly Lorentz invariant.  Moreover, the energy-like dynamical characteristics contain the term that depends on the {\it accelerations} of particles and can, in principle, substitute for the field degrees of freedom (see below).  

As to the (electromagnetic, curvature, etc.) fields themselves, they are uniquely defined via the principal spinor of a {\it shear-free null congruence} of lightlike rays produced by pointlike objects moving along an arbitrary single worldline on the real space-time or even on its complex extension (see, e.g.,~\cite{Newman1,Newman2,Burin1,Burin2,Duplic}). Moreover, at the instants of merging of a pair of particles the fields form a singular null straight line, {\it caustic}, that connects the points of merging and observation and is, perhaps, responsible for the information transmission~\cite{YadPhys}. However, the energy and momentum of the fields, as we have seen, do not explicitly contribute into the structure of conservation laws.    

We have shown that there exist two essentially different classes of polynomial worldlines, the so-called ``spacelike'' or ``timelike'' ones~\footnote{The third, rather remarkable ``lightlike'' class of worldlines corresponding to the case $p=n$ is under study}, and the latter seems to be more suitable for physical interpretation.  For the whole R-C system of particles (or an approximately isolated {\it subsystem} of these), in the timelike case, all the formulas of the canonical relativistic mechanics are valid, including the {\it mass shell equation}  (\ref{restmass}). However, expressions for the momentum or energy of an {\it individual} particle preserve in fact their non-relativistic form and, taken separately, do not satisfy the mass shell constraint. Moreover, velocities 
of individual particles can exceed the light limit and infinitely increase before their merging (annihilation of a pair). However, at {\it late} stages of the evolution, when all the mergings  (``events'') are over, velocities of all the particles decrease and are certainly less than that of the light (Sect.4). 

Anyway, the algebraic dynamics under consideration, though relativistic invariant, strongly differs from the canonical mechanics of Special Relativity. 
On the other hand, it is well known that there are a lot of ``eternal'' problems in the application of the SR-formalism itself (contrary to that of the non-relativistic mechanics).  For example, the energy of fields diverge for the pointlike objects; physical characteristics of particles themselves are, in fact, compared at different proper time instants ( different retardation intervals), etc. In this connection, there were a number of attempts (see, e.g., \cite{Gordeev,Gordeev2,Weert}) to  reformulate the theory in the terms of the unique (laboratory) time. To do this, the authors proved it possible to totally substitute the field degrees of freedom by the infinite set of higher order time derivatives of the coordinates of individual particles. The situation arising in the above presented scheme seems to be rather close to this picture. 

As to the most interesting properties of the developed algebraic dynamics, these are, of course, the {\it ``self-quantization''} of total rest energy of the R-C system of particles (which can take only integer values) and the general recession of particles' ensemble. The latter is moreover accompanied  by formation of pairs and then {\it clusters} of particles at {\it very large} values of the observer's proper time $T\gg T_0$. Qualitatively, the arising picture resembles that of (elastic/inelastic) scattering of a beam of (composite) particles.  One can even speculate about some likeness between the scenario of evolution of the system of clusters and the cosmological evolution of the Universe as a whole. 

As an obvious generalization  of the above presented construction, one can consider the class of worldlines parametrized by {\it rational} functions. Preliminary investigations demonstrate that {\it for a wide class of rational worldlines the whole set of conservation laws will be satisfied} as before. However, the asymptotic behavior in this case will be quite different and even more remarkable from the physical viewpoint.  

Among the obvious physical drawbacks of the theory one can note the effect of {\it `de-isotropization of matter'} which takes place at large values of the ``cosmological'' time $T$: all the particles run away along one or two privileged spatial directions (see item 4 in Sec. 4). Unfortunately, there is also a negligibly small  number of R-particles, at least asymptotically. However, the situation could become much more realistic in the case of an {\it implicitly defined} polynomial worldline~\cite{Khasan,Vestnik} which, in a sense, substitutes for the whole set of ordinary, parametrically defined ones. Corresponding construction (exploiting again the light cone equation) will be considered elsewhere. 

In any case, above revealed purely mathematical properties of the fundamental equation of the (polynomially parametrized) light cone are quite unexpected and seem to be remarkable by themselves. On the other hand,  they explicitly relate to the well-known fundamental properties of physical matter (conservation, quantization, formation of clusters, recession, etc.). We believe that this correspondence is not casual, and a lot of other, physically relevant, ``numerical'' relations  are encoded in the mathematical structures like those examined in the paper.      
\appendix
\section{Structure of Sylvester matrices and conservation laws}

Let us proof the structure (\ref{result1}) of the resultant $D_x=Res[F,H_x,\tau]$ of two polynomials $F(\tau)$ and $H_x=x-X(\tau)$. We start from the simplest case $n=2$; then the two corresponding equations (\ref{terms}) and (\ref{coord}) are of the form
\be{Fn2}
\fl F=A\tau^4 +B\tau^3 + C \tau^2 + (D +2T)\tau +(E-T^2) =0, ~~~(A=\vert \vec a \vert^2\ne 0),
\ee
and 
\be{Xn2}
\fl H_x=a_x \tau^2 + b_x \tau + (e_x-x) = 0, ~~~(a_x \ne 0).
\ee

 The  $6\times 6$ Sylvester matrix $\Sigma_x$ for the above resultant will have the following form:  
\be{sylvest}
\Sigma_x= 
\left(\begin{array}{llllll} 
A & B & C & D +2T & E - T^2 & 0\\  
0 &A & B & C & D +2T & E - T^2 \\
0 & 0 & 0 & a_x & b_x & e_x - x \\ 
0 & 0 &  a_x & b_x & e_x - x & 0 \\ 
0 & a_x & b_x & e_x - x & 0 & 0\\ 
a_x & b_x & e_x - x & 0 & 0 & 0 
\end{array} \right). 
\ee                    
To obtain the resultant, one should compute the {\it determinant} $D_x$ of (\ref{sylvest}) which is obviously a polynomial of the degree $2n=4$ in $x$; we, however, are interested mostly in three leading terms of degrees $4,3,2$ in $x$ written out in (\ref{result1}). Their structure and, in particular, (in)dependence on the time $T$ follow directly from the form of Sylvester matrix (\ref{sylvest}) and completely correspond to that  adduced in the text:
\be{resultmoment}
D_x=\det \Sigma_x = P x^4 + Q_x x^3 + (R_x + E_x T +G_x T^2)x^2 + \dots , 
\ee
where, as it is easy to see, $P = A^2, \dots,  G_x= -2A a_x^2$. Thus, the resultant equation $D_x=0$ leads to the conservation laws of the form represented by (\ref{Moment}) and (\ref{conservsquare}). It is evident (and had been confirmed by the computer algebra computations) that a quite analogous structure of the above determinant-resultant will be observed for any $n$. Specifically, for the leading coefficients $P$ and $G_x$ one obtains the expressions (\ref{coeffX}) $\Box$ 

Let us examine now the structure of resultant $R_x:=Res[F,N_x,\tau]$ of the two polynomials $F(\tau)$ and $N_x$ (where the latter is represented by (\ref{polynomang}) and 
 related to the $x$-component of the angular momentum (\ref{componang})). For this case the Sylvester matrix has the form essentially different from (\ref{sylvest}). Specifically, for $n=2$ the principal light cone equation is represented by (\ref{Fn2}) while the polynomial equation (\ref{polynomang}), on account of the defining equations of the worldline (\ref{polynomsXi}), takes the form
\be{Mn2}
\fl N_x= M_x (4A\tau^3+3B\tau^2+2C\tau + D+2T) - (T-\tau) (S_x \tau^2 + L_x \tau + K_x) = 0, 
\ee 
where, in particular, 
\be{coeff1}
\fl S_x = 2(b_y a_z - a_y b_z)
\ee 
and all $S_x,L_x,K_x$ are independent on $T$.                                                                                                                                                                                                                             
                                  
Designating now for short $\tilde D = D+2T,~~\tilde E = E-T^2$ and omitting for a time 
the component index $x$, we are able to write out   
the corresponding $7\times 7$ Sylvester matrix of the resultant $R=Res[F,N,\tau]$:
\be{sylvest2}
\fl \left( \small
\begin{array}{lllllll}
A & B & C & \tilde D  & \tilde E & 0 & 0 \\
0 & A & B & C & \tilde D & \tilde E & 0 \\
0 & 0 & A & B & C & \tilde D & \tilde E\\ 
0 & 0 & 0 & 4A M + S & 3B M + U & 2C M +V & \tilde D M + W \\
0 & 0 & 4A M + S & 3B M +U & 2C M +V & \tilde D M +W & 0\\
0 & 4A M + S & 3B M + U & 2C M +V & \tilde D M + W & 0 & 0\\
4A M + S & 3B M + U & 2C M +V & \tilde D M +W & 0 & 0 & 0
\end{array}\right)
\ee
where $S(=S_x)=const$ while  $U(=U_x)=L_x-T S_x,~V(=V_x)=K_x-T L_x,~W(=W_x)=-T K_x$ are linear functions of $T$. 

Now it is easy to see that the first two leading terms in the determinant of (\ref{sylvest2}) of degrees $4$ and $3$ in $M$ do not involve any of time-dependent coefficients $U,V,W$. The leading term of degree $4$ does not also contain $S$ and, whether all these coefficients are set zero, the determinant of  (\ref{sylvest2})  becomes equal to $M^4 Res[F,F^\p,\tau]\equiv M^4 \Delta(T)$, that is, proportional to the {\it discriminant} $\Delta(T)$ of the principal light cone equation $F=0$. The next term of degree $3$ contains the constant $S$, is also proportional to the discriminant $\Delta(T)$ and  equal exactly to $M^3 \Delta(T) (S/ A)$ so that one has finally:  
\be{angresnew}
R_{(n=2)}=\Delta(T) M^4 + \alpha \Delta (T) M^3 + ... =0,~~~ (\alpha = S/A). 
\ee
It is quite easy to observe now that the generalization of the above considerations to any $n>2$ leads to the structure of resultant $R_x=Res[F,N_x,\tau]$  similar to (\ref{angresnew}), namely (we can restore now the component index $x$):
\be{angresnewN}
R_x=\Delta(T) M_x^{2n} + \alpha_x \Delta (T) M_x^{2n-1} + ... =0,~~~ \alpha_x = S_x/A,  
\ee
that is, to the structure completely identical to that of (\ref{angVieta}) presented in the text. The quantity $-\alpha_x$ which determines the ($x$-component of) the total vector of angular momentum is in fact equal to the ratio of the leading coefficients $S_x(=S)$ and $A$ of corresponding polynomials. Substituting their expressions (\ref{coeff1}) and (\ref{Fn2}) through the coefficients of defining polynomials (\ref{polynomsXi}) one obtains finally:
\be{sumangX}
\sum (M_x)_i =-\alpha_x =2\frac{a_yb_z-b_ya_z}{a_x^2+a_y^2+a_z^2}, 
\ee
and, of course, analogous formulas for two other components. Thus, we proof the universal formula (\ref{explicang}) for the vector $\vec M$ of total angular momentum of the system of R-C particles. It is noteworthy that the expression for $\vec M$ remains the same for any degree $n$ of the defining polynomials (\ref{polynomsXi}). 
 $\Box$

\appendix
\setcounter{section}{1}
\section{Dynamical characteristics for the examplified worldline}

In Sect.4 we presented a number of figures illustrating the collective dynamics of $2p=48$ roots-particles for the timelike ($p>n$) worldline with $p=24,n=20$. Specifically, we accepted the following (randomly selected) form of the  polynomials (\ref{timelike})  parameterizing the worldline: 
\be{examplpol}
\begin{array}{llll}
S= \tau^{24}+5\tau^{23}-2\tau^{22}+21\tau^{19}-\tau^7+5\tau-8, \\
X= \tau^{20}-11\tau^{19}-13\tau^{18}-3\tau+5, \\
Y= 2\tau^{20}+17\tau^{19}+7\tau^{18}+2\tau^2-1, \\
Z= -3\tau^{20}+13\tau^{19}-9\tau^{17}+7\tau^4+12. 
\end{array}
\ee
Substituting (\ref{examplpol}) into the light cone equation (\ref{timelikecone}), taking the resultant $Res[F,G,\tau]$ of its left-hand-side polynomial with that of $G$ represented by (\ref{EqForS}) and equating the resultant to zero, one obtains the following polynomial equation for determination the dependence $s=s(T)$:
\be{results2}
s^{48}-(48 T + \delta)s^{47}+(1128 T^2 +\varphi T+\psi)s^{46}+\dots =0,  
\ee
where $\delta,\varphi,\psi$ are some great integer-valued real numbers (their exact values being irrelevant in what follows).  
From (\ref{results2}), for the sums of all 48 roots and their squares it follows immediately:
\be{sum_s}
\sum_i s_i = (48 T+\delta),~\sum_i s_i^2 = (48 T + \delta)^2 -2 \cdot (1128 T^2+\varphi T +\psi),
\ee
so that after appropriate differentiations w.r.t. $T$ we obtain the following conserved values for the rest energy (rest mass) of the system of R-C particles under consideration:
\be{restmass}
E=\sum_i {\dot s}_i = 48 =2p, ~~W_0 = \sum_i {\dot s}_i^2 + {\ddot s}_i s_i = 48 \equiv E,
\ee
in full agreement with the general formulas (\ref{4-energy-conserv},\ref{0-virial}) or (\ref{identity}). Of course, it is easy to assure that the total momentum $\vec P$ represented by (\ref{timelikemomentum}) as well as the corresponding quantities $W_a$ as given by (\ref{a-virial}) turn to zero in the assumed rest frame of the observer. 

Consider now the skew tensor of angular momentum (\ref{tensorang}). All its six components are nonzero and  conserved but their direct calculation by means of the procedure described in Sect.2 and Appendix A is rather cumbersome since their values turn out to be {\it unexpectedly great}. Nonetheless, with the help of the computer algebra system {\it Mathematica}, we find in the end: 
\begin{equation}
\fl \begin{array}{ll}
M_x=5585145722802,~M_y=11642674433710,~M_z=316030077916,\\
K_x=2277192433512854,~K_y=-799390472683014,~K_z=-4701753623402064,
\end{array}
\end{equation}
where the components $\{M_a\}$ are defined by (\ref{vectangmom}) while the quantities $\{K_a\}$ correspond to the components $M_{[0a]}$ of the angular momentum tensor. The two $SO(3)$-invariants are, respectively, 
\be{angSO(3)}
\begin{array}{ll}
{\vec M}^2:=M_x^2+M_y^2+M_z^2 = 16684559574445537237998360,\\
{\vec K}^2=K_x^2+K_y^2+K_z^2=27931117642239004403999706809608, 
\end{array}
\ee
while the conserved Lorentz ($SO(3,1)$) scalar is 
\be{andSO(3,1)}
\sqrt{{\vec K}^2-{\vec M}^2} =\dots \approx 2.793111763\cdot 10^{31}.
\ee
Of course, these exact results have been checked by direct numerical calculations of the sums of roots and respective dynamical quantities of these for different values of the observer's time $T$. 

 The discriminant polynomial equation $\Delta(T)=0$ possesses 8 real roots defining the instants $T_{merge}$ of the annihilation/creation events. Specifically, at the interval $t_1 <T< t_2,~~t_1\simeq -18.8923,~t_2\simeq -18.3244 $ the light cone equation has 8 (maximal possible number for the considered example) real and 20 pairs of complex conjugate roots defining 8 R-particles and 20 (composite) C-particles, respectively. At the moment $T\simeq t_1$ one has a creation of a pair of two R-particles (precisely, of a pair particle/antiparticle) while at the moment $T\simeq t_2$ there is a merging of two R-particles accompanying by the formation of one C-particle, and so on. The first event (creation of the first R-pair) takes place at the great negative value $T\simeq -0.7548374411\cdot 10^{16}$. The last event occurs at $T\simeq 4.434361164$, and after this moment the number of R- and C-particles remain invariable (four R-particles and 22 C-particles). 

The critical values $T_1=\simeq 2.7\cdot 10^3$ and $T_0\simeq 6.0\cdot 10^{16}$ as defined in the text (Sect.4) mark the intervals at which the particles are separated ($0<T \le T_1$), join into pairs ($T_1<T \le T_0$) or start to gather into clusters ($T\gg T_0$), see figs.2 and 3, respectively.

\end{document}